\documentclass[prd,twocolumn,showpacs]{revtex4-1}
\usepackage{epsfig,amsmath,amssymb,mathrsfs,mathtools,amscd}
\usepackage[linkcolor=red]{hyperref}
\usepackage{enumerate}
\newtheorem{lemma}{Lemma}

\newtheorem{proposition}{Proposition}

\newtheorem{corollary}{Corollary}

\newtheorem{definition}{Definition}

\makeatletter
\renewcommand*{\@opargbegintheorem}[3]{\trivlist
      \item[\hskip \labelsep{\bfseries #1\ #2}] \textbf{(#3)}\ \itshape}
\makeatother

\usepackage{comment}
\usepackage{cleveref}
\usepackage{float}
\usepackage{soul}
\usepackage{amsmath,amsfonts,amssymb}
\def\kmom{\mathrm{k}}
\def\Diff{\mathrm{Diff}}

\mathchardef\minus="002D

\def\Proof{\medskip\par\noindent{\bf Proof. }}
\def\qed{$\,\blacksquare$\par}
\def\<{\langle}
\def\>{\rangle}
 \def\ket#1{| #1 \rangle}

\def\ketbra#1#2{| #1 \rangle \langle#2 |}

\begin{document}

\title{Symmetries of the Dirac quantum walk and emergence of the de Sitter group}

\author{Luca \surname{Apadula}}
\email[]{luca.apadula01@ateneopv.it} 
\affiliation{University of Vienna, Boltzmanngasse 5, 1090 Vienna, Austria }\affiliation{Institute for Quantum Optics and Quantum Information (IQOQI), Austrian Academy of Sciences, Boltzmanngasse 3, 1090 Vienna, Austria}
\affiliation{Dipartimento di
  Fisica, Universit\`a di Pavia, via Bassi 6, 27100 Pavia, Italy}

\author{Alessandro \surname{Bisio}}
\email[]{alessandro.bisio@unipv.it} \affiliation{Dipartimento di
  Fisica, Universit\`a di Pavia, via Bassi 6, 27100 Pavia, Italy}
\affiliation{Istituto Nazionale di Fisica Nucleare, Sezione di Pavia,
  Italy}

\author{Giacomo Mauro \surname{D'Ariano}}
\email[]{dariano@unipv.it} \affiliation{Dipartimento di
  Fisica, Universit\`a di Pavia, via Bassi 6, 27100 Pavia, Italy}
\affiliation{Istituto Nazionale di Fisica Nucleare, Sezione di Pavia,
  Italy}

\author{Paolo \surname{Perinotti}}\email[]{paolo.perinotti@unipv.it}
\affiliation{Dipartimento di Fisica, Universit\`a di Pavia, via Bassi 6, 27100 Pavia, Italy}
\affiliation{Istituto Nazionale
  di Fisica Nucleare, Sezione di Pavia, Italy}

\begin{abstract}
  A quantum walk describes the discrete unitary evolution of a quantum
  particle on a discrete graph. Some quantum walks, referred to as the Weyl and
  Dirac quantum walks, provide a description of the free evolution of
  relativistic quantum fields in a regime where the wave-vectors
  involved in the particle state are small. The clash between the
  intrinsic discreteness of quantum walks and the symmetries of special
  relativity can be resolved by rethinking
  the notion of a change of inertial reference frame. We give here a
  definition of the latter that avoids a pre-defined space-time
  geometry, 
  in terms of a
  change of values of the constants of motion that leaves the walk
  operator unchanged. Starting from the family of 1+1 dimensional Dirac quantum walks with all possible 
  values of the mass parameter, we introduce a unique walk encompassing the latter
  as an extra degree of freedom, 
  and we derive its group of changes of inertial frames.  
  This symmetry
  group contains a non linear realization of $SO^+(2,1)
  \ltimes \mathbb{R}^3$;
  since one of the two space-like dimensions does not
  correspond to an actual spatial degree of freedom but rather the
  mass, we interpret it as a 2+1 dimensional de-Sitter group. 
  This group group contains also a non-linear realization of the proper orthochronous Poincar\'e
  group $SO^+(1,1) \ltimes \mathbb{R}^2$ in
  1+1 dimension, as the ones considered within the frameork of doubly
  special relativity, which  recovers the usual relativistic symmetry of the
  Dirac Equation in the
  limit of small wave-vectors and masses.
  Surprisingly, if one considers the Dirac
  walk with a fixed value of the mass parameter, the group of allowed
  changes of reference frame does not have a consistent interpretation
  in the relativistic limit of small wave-vectors.
\end{abstract}
\maketitle  

\section{Introduction}

The reconciliation of quantum theory with general relativity is one of
the most ambitious goals of contemporary physics, and counts a wealth
of approaches based on radically different standpoints. One of the
ideas behind some of the relatively recent approaches is the proposal
that space-time might be a derived notion instead of a primitive one,
thus emerging from some non-geometric underlying structure
\cite{bombelli1987space,jacobson1995thermodynamics,
  PhysRevD.77.104029,PhysRevD.81.104032,Verlinde:2011ab,green_schwarz_witten_2012}.

The reconstruction of free quantum field theory through principles controlling the processing of information carried by elementary quantum systems 
\cite{PhysRevA.90.062106,Bisio:2016aa,bisio2013dirac,BDT15,PhysRevA.97.032132}
constitutes one of the promising approaches to emergent physical laws. 
A characteristic trait of this approach is that the starting structure 
is a quantum cellular automaton \cite{Schumacher:2004ab}, i.e.~a discrete array of 
memory cells, governed by an update rule that 
acts in a discrete sequence of evolution steps. A similar model 
for elementary physical processes was the subject of Feynman's 
pioneering proposal of a universal quantum simulator \cite{feynman1982simulating}.

Some of the approaches to the dynamics of quantum fields based on quantum 
cellular automata and their simplified description through quantum walks implicitly 
assume a pre-defined geometry, translated into the properties of the quantum gates
producing the evolution of the cellular automaton \cite{1367-2630-16-9-093007,Arrighi2016,
Raynal:2017aa,PhysRevA.97.042131,DiMolfetta2015}

One of the key features of the approach initiated by some of the present 
authors is the fact that space-time is not a primitive notion in this framework, 
while geometry emerges only in the presence of  evolving systems---quantum fields---in terms of the symmetries of their dynamics.

The intimate discreteness of cellular automata appears at odds with 
the symmetries of known physical laws, in particular the Poincar\'e 
group of special relativity. It was already proved in 
Refs.~\cite{BBDPT15, Bisio20150232,Bisio:2016ab} that for the Weyl automata the
Poincar\'e symmetry can be recovered by generalizing the relativity
principle, defining changes of inertial frames as those changes of
representation of the cellular automaton---in terms of the values of its
constants of motion---that preserve the update rule. Such a notion is suitable 
to the study of dynamical symmetries, without the need of resorting to a 
space-time background. 

The above mentioned result represents a proof of principle that a
discrete quantum dynamics is consistent with the symmetries of
classical space-time. The Poincar\'e group acts on the space of wave
vectors through a {\em realization}---in the present case a group of
diffeomorphisms---instead of the usual linear representations of
quantum field theory. The non-linearity deformation of the Poincar\'e
symmetry is the distinctive feature of doubly-special relativity (DSR)
models,
\cite{amelino2001testable,magueijo2003generalized,PhysRevD.84.084010},
which consider theories with two observer-independent scales, the
speed of light and the Planck energy.
Recently, experimental tests of violation of Lorentz symmetry
have been proposed
\cite{Moyer:2012ws,Hogan:2012ik,pikovski2011probing}.
In particular,
observation of deep space gamma-ray bursts
can be sensitive to the vacuum dispersive behaviour
\cite{Amelino-Camelia:1998aa,Abdo:2009aa2,Vasileiou:2015aa,Amelino-Camelia:2017aa}.

A partial classification of the full symmetry group of the Weyl
automaton in 3+1 dimensions was derived in Ref.~\cite{Bisio2017}.  In
the present paper we provide an extension of the analysis to the case
of the Dirac automaton in 1+1 dimensions, namely an automaton where
the extra parameter representing mass plays an important
dynamical role. If one considers a Dirac automaton with a fixed value
of the mass parameter, one finds a symmetry group that is isomorphic
to $SO^+(1,1)\rtimes\mathbb Z_2$, namely the Lorentz group in 1+1
dimensions. However, the analysis of the action of such group in terms
of its action in the limit of small wave-vectors is inconsistent with
the identification of the wave-vector with momentum.
Therefore, we introduce a quantum walk in which
the mass is an extra degree of freedom, on the same footing
as the wave-vector. 


The symmetry group is proved to be the semidirect product of three
groups.
The first one is
the additive group of  smooth functions from the invariant zone to the
complex numbers. The second one is a group of diffeomorphisms
which act as  nonlinear dilations of the quantum walk mass shell.
The third group is a non linear realization of  $SO^+(2,1)$.
We observe that the symmetry
  group contains a non linear realization of $ISO^+(1,2) = SO^+(2,1)
  \ltimes \mathbb{R}^3$ which is
  interpreted as a variation of the de Sitter group,
  in the limiting flat case of infinite cosmological constant.
The reason for this is that the extra dimension emerging in our case is not a spatial
one, but it is associated with the variable mass parameter.  This result introduces an inspiring relation between 
the symmetries of the massive quantum field and those of the emerging space-time geometry.  Notice also that the  classical-particle interpretation of the conjugated variable of the rest mass is that of proper time \cite{doi:10.1063/1.1665400}. 

Within the subgroup $SO^+(1,2)\ltimes\mathbb{R}^3$
we also have a non linear representation of the Poincar\'e
group $SO^+(1,1)\ltimes\mathbb{R}^2$ in 1+1 dimension
which, contrarily to the fixed mass case,
consistently recovers the usual (linear) relativistic symmetry
in the limit of small wave-vectors. Therefore,
the Dirac quantum walk naturally provides a microscopic dynamical model of
doubly special relativity.


The paper is organised as follows:
Section~\ref{sec:one-dimens-dirac} begins with a 
review of basic notions of quantum walks on Cayley graphs and of the one dimensional
Dirac quantum walk. Then, in section~\ref{sec:variable-mass}
we introduce the one dimensional Dirac quantum walk with variable
mass, whose eigenvalue equation is studied in
Section~\ref{sec:study-eigenv-equat}.
In Section \ref{sec:change-inert-frame} we define a notion of change
of inertial frame which does not rely on a symmetry of a background
spacetime. We then characterize the group of changes of inertial
frames of the Dirac walk with variable mass and we show that it
consists in a non-linear realization of a semidirect product of the
Poincar\'{e} group and the group of dilations.

\section{The one dimensional Dirac quantum walk}
\label{sec:one-dimens-dirac}

\begin{figure}[t]
 \includegraphics[width=0.8\columnwidth]{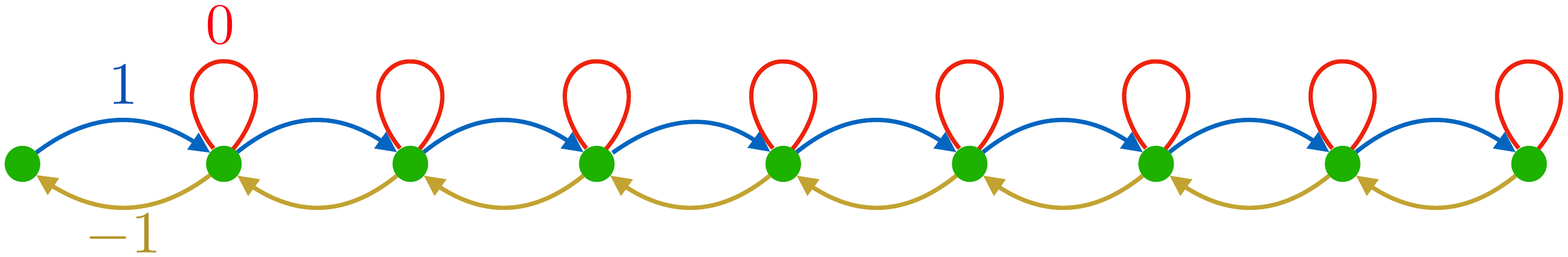}
 
 \includegraphics[width=0.5\columnwidth]{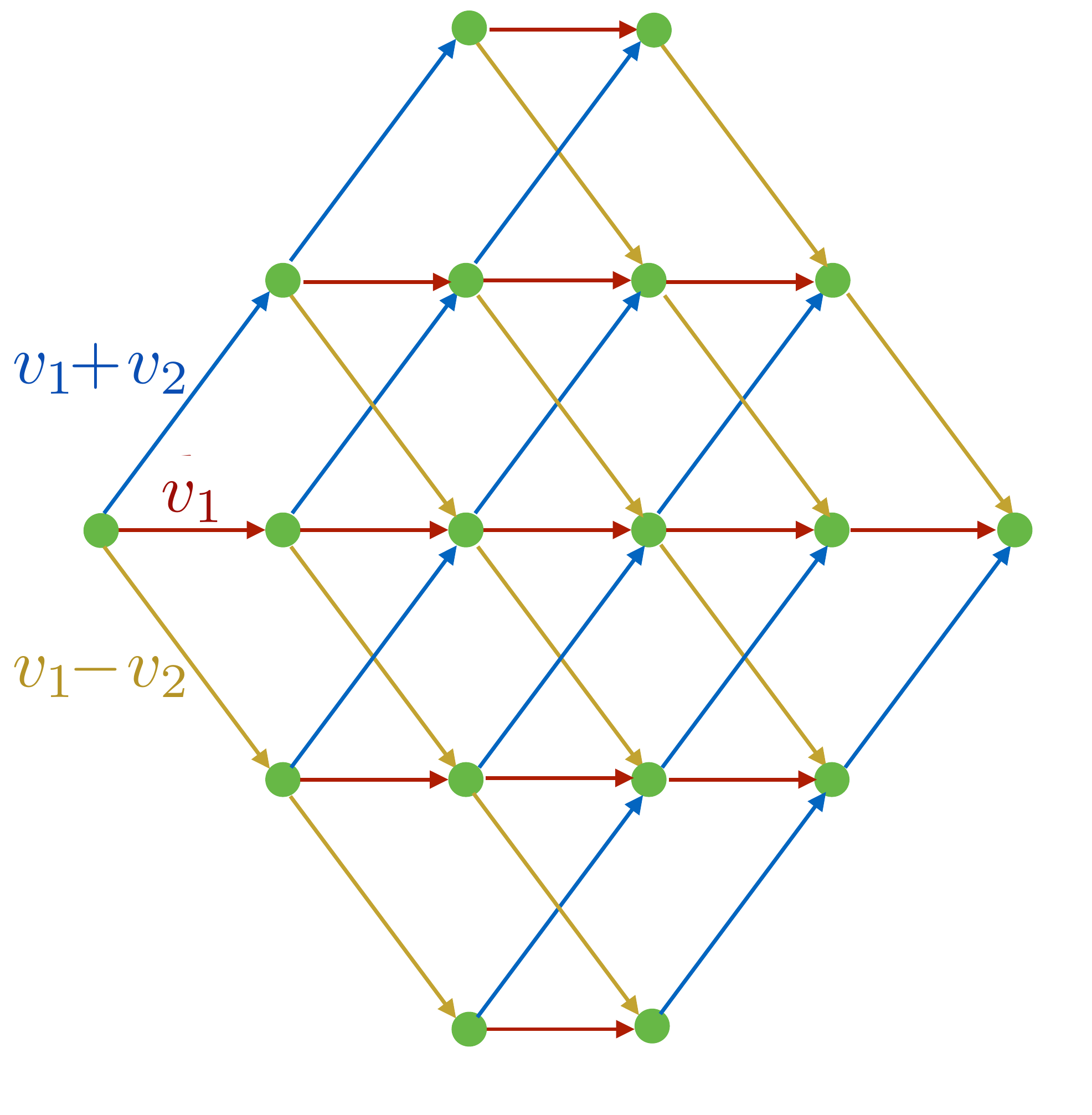}
 \caption{\emph{Top}: Cayley graph of the group $\mathbb{Z}$, where the red loop arrow represents the identity $\{0\}$, and the blue/yellow arrow refer to right/left translation, namely $\{1,-1\}$.\\
\emph{Bottom}: Cayley graph of the group $\mathbb{Z}_2$, where the red arrow is associated to the generator $g_1$, while the blue and yellow arrows refer to the generators $g_1+g_2$, $g_1-g_2$ respectively. }
   \label{Cayley}
 \end{figure}

A discrete time quantum walk \cite{AB01,portugal2013quantum} describes
the unitary evolution of a particle with $s$ internal degrees of
freedom (usually called \emph{coin space}) on a lattice $\Gamma$.  In the case of interest \cite{PhysRevA.90.062106}
the lattice $\Gamma$ is the \emph{Cayley graph} of a finitely
generated group $G$, i.e.  $\Gamma(G,S_+)$ is the edge-colored directed
graph having vertex set $G$, and edge set
$\{ (x,xh), x \in G, h \in S=S_+\cup S_+^{-1} \}$ ($S_+$ is a set of generators of $G$), and a color assigned to each
generator $h \in S_+$.  Usually, an edge
which corresponds to a generator $g$ such that $h^2 = e $ ($e$ is the
identity of $G$) is represented as undirected (as the green arrow of
the first graph in Fig.~\eqref{Cayley}).  Clearly, each Cayley graph
corresponds to a presentation of the group $G$, where relators are
just closed paths over the graph.  Within this framework, a
discrete-time quantum walk on a Cayley graph $\Gamma(G,S_+)$ with an
$s$-dimensional coin system ($s\geq 1$) is a unitary evolution on the
Hilbert space $\ell^2(G)\otimes \mathbb{C}^s $ of the following kind
\begin{align*}
  \begin{aligned}
  A \coloneqq \sum_{h\in S} T_{h} \otimes A_h , \\
  0\neq A_h \in M_s(\mathbb{C}),\\
  T_h|x\>\coloneqq |xh^{-1}\>,  
  \end{aligned}
\end{align*}
where,  for any $g\in G$, $T_g$
is the right regular representation of $G$ on
$ \ell^2(G)$ and $ \{ \ket{x}, x \in G \}$
is an orthonormal basis of $\ell^2(G)$.

The one dimensional Dirac quantum walk is a quantum walk on the Cayley
graph $\Gamma(\mathbb{Z}, \{ 0, 1 \} )$ (see Fig.~\eqref{Cayley}) of the group
$\mathbb{Z}$ with coin space $\mathbb{C}^2$ (the particle has two
internal degrees of freedom).  The evolution is the following unitary
operator on $\ell^2(\mathbb{Z}) \otimes \mathbb{C}^2$:
 \begin{align}\label{DiracQW}
   \begin{aligned}
{A(\mu)}=\begin{pmatrix}
  T\cos\mu  &-i I\sin\mu\\
  -i I\sin\mu & T^\dag\cos\mu 
\end{pmatrix}, \\
\ket{\psi} = \sum_{s=L,R}\sum_{x\in \mathbb{Z}} \psi(s,x,\mu) \ket{x}     \ket{s}\\
     \ket{R} =
     \left(
     \begin{array}{c}
       1 \\
       0
       \end{array}
     \right),
     \quad
     \ket{L} =
     \left(
     \begin{array}{c}
       0 \\
       1
       \end{array}
     \right)      
   \end{aligned}
\end{align}
where $I=T_0$, $T := T_{1}$,\,            $ T_1 \ket{x} = \ket{x+1}$
and  $T^\dag = T_{-1}=T^{-1}$.
Since $A(\mu)$ commutes with the translation operator $T\otimes I_2$,
we may represent $A(\mu)$ using the Fourier basis
$\ket{\kmom} = \frac{1}{\sqrt{2 \pi}} \sum_{x \in \mathbb{Z}}e^{i
  \kmom x } \ket{x}$
and we obtain
\begin{align}
  \label{eq:QCAfourier}
  \begin{aligned}
      A(\mu) = \int_{-\pi} ^{\pi}   d\kmom  \, \ketbra{\kmom}{\kmom}\otimes\tilde A(\mu, \kmom)   \\
    \tilde A(\mu, \kmom)=
\begin{pmatrix}
\cos{\mu} e^{-i\kmom} & i \sin{\mu} \\
i \sin{\mu} &  \cos{\mu} e^{i\kmom}
\end{pmatrix}.
  \end{aligned}
\end{align}
In the limit $\kmom,\mu \to 0 $ the Dirac Quantum Walk
recovers the dynamics of the one dimensional Dirac equation, where $\kmom$
and $\mu$ are interpreted as momentum and mass of the particle respectively.
\subsection{Variable mass}
\label{sec:variable-mass}
As we already announced in the introduction, and will be shown in
the following section, the symmetry group of
the Dirac walk cannot recover the relativistic Lorentz
symmetry.
This obstruction can be overcome by considering the mass no longer as
a fixed parameter, but rather as an
additional degree of freedom, as follows
\begin{align}
  \begin{aligned}
    \label{eq:variablemass}
  A := \int_{-\pi}^{\pi}        d\mu  A(\mu)  \otimes
  \ketbra{\mu}{\mu} \\
  \ket{\mu} :=
  \frac{1}{\sqrt{2\pi}} \sum_{\tau \in \mathbb{Z}} e^{i \mu \tau} \ket{\tau}
  \end{aligned}
\end{align}
where $\ket{\tau}$ is an orthonormal basis of $\ell^2(\mathbb{Z}) $. The discrete nature of the variable conjugated to the mass agrees with the discreteness of time in the quantum walk, being $\tau$ interpreted as the proper time of the classical particle 
\cite{doi:10.1063/1.1665400}
It is easy to realize that $ A$ is a Quantum walk on a Cayley graph 
of $\mathbb{Z}^2$.
Indeed, from Eq.~\eqref{eq:variablemass} we have 
\begin{align}
  \begin{aligned}
    \label{eq:variablemass2}
   &  A := \int_{B}
        d\kmom \, d\mu \,
   \tilde A(\mu,\kmom)  \otimes \ketbra{\mu,\kmom}{\mu,\kmom}\\
 & \tilde A(\mu,\kmom) =
  \frac{1}{2}
  \begin{pmatrix}
(e^{i\mu}+e^{-i\mu})e^{-i\kmom}& (e^{i\mu}-e^{-i\mu})\\
e^{i\mu}-e^{-i\mu}& (e^{i\mu}+e^{-i\mu})e^{i\kmom}
\end{pmatrix}
\\
&  \ket{\mu,\kmom} := \ket{\mu}\ket{\kmom},\\
& B:=   (-\pi, \pi ] \times (-\pi, \pi ].    
  \end{aligned}
\end{align}
which in the $\ket{x}\ket{\tau}$ basis reads
\begin{align}
  \begin{aligned}
    \label{eq:variablemass3}
    &A = \frac{1}{2} \begin{pmatrix}
(T^\dag+T)S& T^\dag-T\\
T^\dag-T& (T^\dag+T)S^\dag
\end{pmatrix},\\
&T \ket{\tau} = \ket{\tau+1}.  
  \end{aligned}
\end{align}
In the right regular representation of $\mathbb{Z}^2$, with basis 
$\ket{x}\ket{\tau}$,  $T$ and $S$ represent the generator $g_1 := (0,-1)$
and $g_2:= (-1,0)$ respectively.
Therefore $A$ is a quantum walk on the Cayley graph
$\Gamma(\mathbb{Z}^2, \{ \pm g_1, \pm (g_1+g_2),  \pm (g_1- g_2)\} )$
(see Fig.~\eqref{Cayley}).

It is worth noticing  that the previous construction depends on the
choice of parametrisation for the mass term in
Equation~\eqref{DiracQW} (for example, the change of variables $\mu' = \sin\mu$
would not have led to a Quantum walk in the conjugate variables).

\subsection{Study of the eigenvalue equation}
\label{sec:study-eigenv-equat}
\begin{figure}[t]
 \includegraphics[width=0.8\columnwidth]{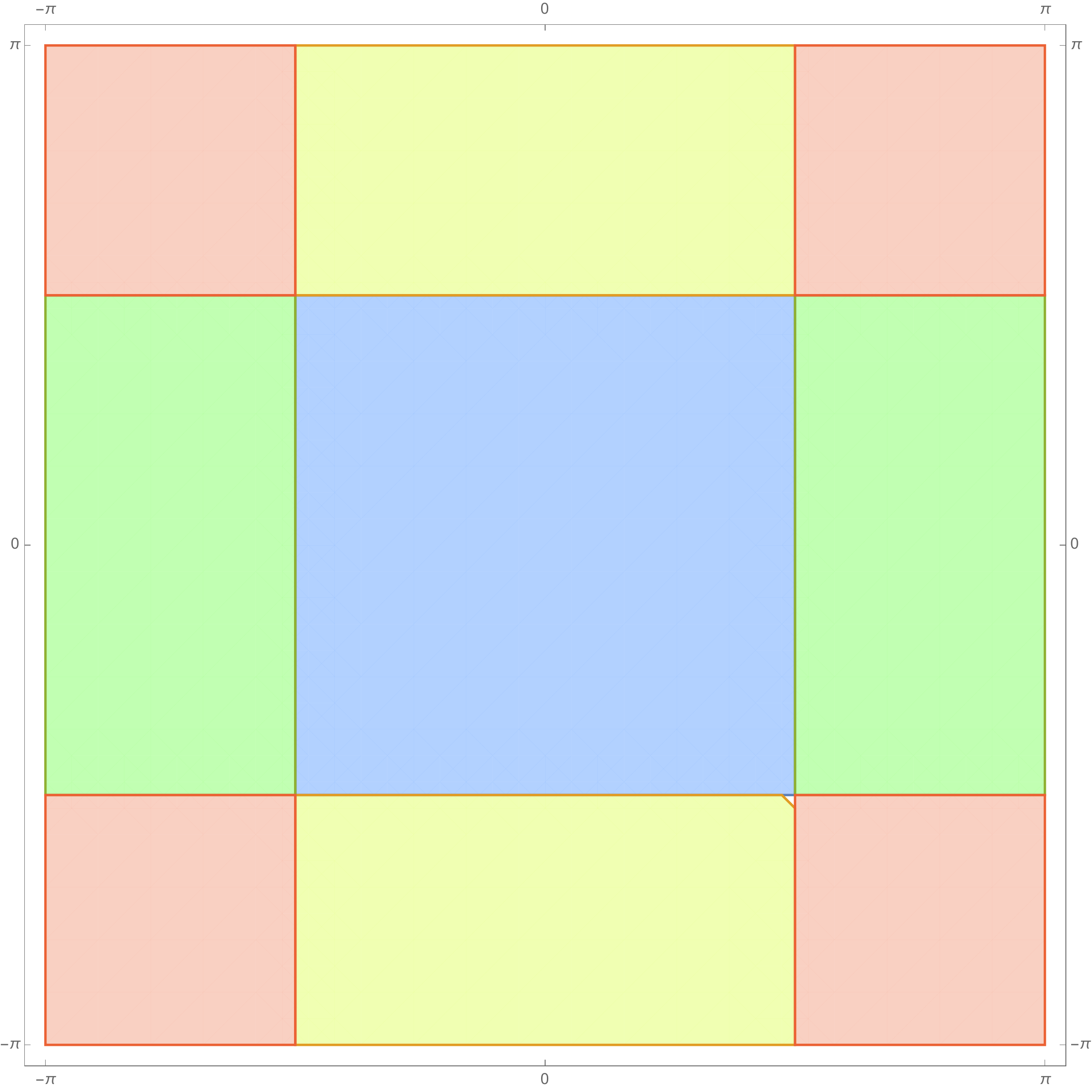}
 \caption{Brillouin zone: the blue region refers to $B_0$, 
 while the other colored regions correspond to $B_i$, and are obtained simply translating by the vector $(a,b)$
 with $a,b\in\{0,\pi/2\}$.}
   \label{Brillouin}
 \end{figure}

Let us consider the eigenvalue equation for the
Dirac Quantum Walk with variable mass.
From Equation~\eqref{eq:variablemass2}
we have

\begin{align}
\label{egv_omega}
  &A(\mu, \kmom)\psi(\kmom,\mu)=e^{i\omega(\kmom,\mu)}\psi(\kmom,\mu), \\
  &\psi(\kmom,\mu) =
  \begin{pmatrix}
    \psi(R,\kmom,\mu) \\
    \psi(L,\kmom,\mu)
  \end{pmatrix}.
  \nonumber
\end{align}
which can be rewritten as 
\begin{align}
 & (\cos\mu\cos\kmom-\cos\omega)\psi(\kmom,\mu)=0\label{eq:bbbanale} \\
 & (\cos\mu\sin\kmom\,\sigma_3-
    \sin\mu\,\sigma_1+\sin\omega\, I)
   \psi(\kmom,\mu)=0.
   \label{eq:projectorrank1}
\end{align}
From the first equation we get the expression for the eigenvalue,
namely $\omega=\arccos(\cos\mu\cos\kmom)$,
while multiplying the second equation by $\sigma_2$ we obtain
\begin{align}\label{eq:eigenvaleq1}
\left(\cos\mu\sin\kmom\,i\sigma_1+\sin\mu\, i\sigma_3+\sin\omega\,\sigma_2\right)\psi(\kmom,\mu)=0.
\end{align} 
We notice that the set $\{\sigma_2,i\sigma_1,i\sigma_3\}$
provides a representation of the generators of the Clifford algebra
$\mathsf{C}\ell _{1,2}(\mathbb{R})$.
Indeed, by renaming the elements of
the set as $\{\tau_1,\tau_2,\tau_3\}$,
the following relations are
satisfied
\begin{equation}\nonumber
\{\tau_i,\tau_j\}=2\eta_{ij},
\end{equation}
where $\eta_{ij} $ denotes the Minkowski metric tensor with signature
$(+,-,-)$. Hence, we can rewrite
equation \eqref{eq:eigenvaleq1} in the
relativistic notation
\begin{align}
\label{egv}
      &n_{\mu}(\kmom,\mu)\tau^{\mu}\psi(\kmom,\mu)=0, \\
&n:=(\sin\omega,\cos\mu\sin\kmom,\sin\mu), \nonumber \\
&\tau:=(\sigma_2,-i\sigma_1,-i\sigma_3). \nonumber
\end{align}
Furthermore, if equation \eqref{egv} holds,
we have
\begin{align}
\label{eq:massshell}
&n_{\nu}(\kmom,\mu)n^{\nu}(\kmom,\mu)=0,
\end{align}
and consequently Eq.~\eqref{eq:bbbanale} is trivially satisfied,
i.e.~$\omega(\kmom,\mu)=\arccos(\cos\mu\cos\kmom)$.
Now, let us analyze the map
\begin{align}
  \label{eq:mapsn}
&  \begin{aligned}
  \bar{n}(\kmom,\mu) : B &\to \mathbb{R}^2 \\
  (\kmom, \mu ) &\mapsto (\cos\mu \sin\kmom , \sin\mu),    
  \end{aligned}
\end{align}
if we compute the norm of the considered map, we have
\begin{align}\label{1disc}
\Vert \bar n(\kmom,\mu)\Vert^2=\sin^2\kmom\cos^2\mu+\sin^2\mu\leq1,
\end{align}
which implies that the Brillouin zone is mapped in the  unit disc in $\mathbb{R}^2$.
Clearly, $\bar{n}$ is smooth and analytic.
The Jacobian of $\bar{n}$ is
\begin{align}\nonumber
J_{\bar{n}}(\kmom,\mu)=\text{det}(\partial_i n_j)=\cos^2\mu\cos\kmom,
\end{align}
and the map results singular for $\kmom={\pi}/{2}+m\pi$ and
$\mu={\pi}/{2}+m\pi$, with $m\in\mathbb{N}$. Let us define the
following regions $B_i \subset B$
\begin{align}
  \label{eq:regionsBi}
  \begin{aligned}
B_0&:=\{(\kmom,\mu) | \kmom \in (-\tfrac{\pi}{2}, \tfrac{\pi}{2} ) , \mu \in (-\tfrac{\pi}{2}, \tfrac{\pi}{2} )  \},\\
B_1&:=B_0 + (\tfrac{\pi}{2},0),\\
B_2&:=B_0 + (0,\tfrac{\pi}{2}),\\
B_3&:=B_0 + (\tfrac{\pi}{2}, \tfrac{\pi}{2} ),
  \end{aligned}
\end{align}
where $B_0 + (a,b)$ denotes the translation of the set $B_0$ by the
vector $(a,b)$ (see Fig.~\ref{Brillouin}). Denoting by $\bar n\vert_{B_0}$ the map $\bar n$ restricted to
the region $B_0$, and referring to Eq.~\eqref{1disc}, 
it is easy to note that $\bar n\vert_{B_0}$ is an analytic diffeomorphism between $B_0$ 
and the open unit disc in $\mathbb{R}^2$. Then, thanks to the periodicity of the map
$\bar n$, the property of being an analytic diffeomorphism straightforwardly holds for 
$\bar n\vert_{B_i},\,\,\forall i\in\{0,1,2,3\}$, $\bar n\vert_{B_i}$ denoting the restriction of $\bar n$ to the region $B_i$.

\begin{figure*}[t]
\includegraphics[width=.32\textwidth]{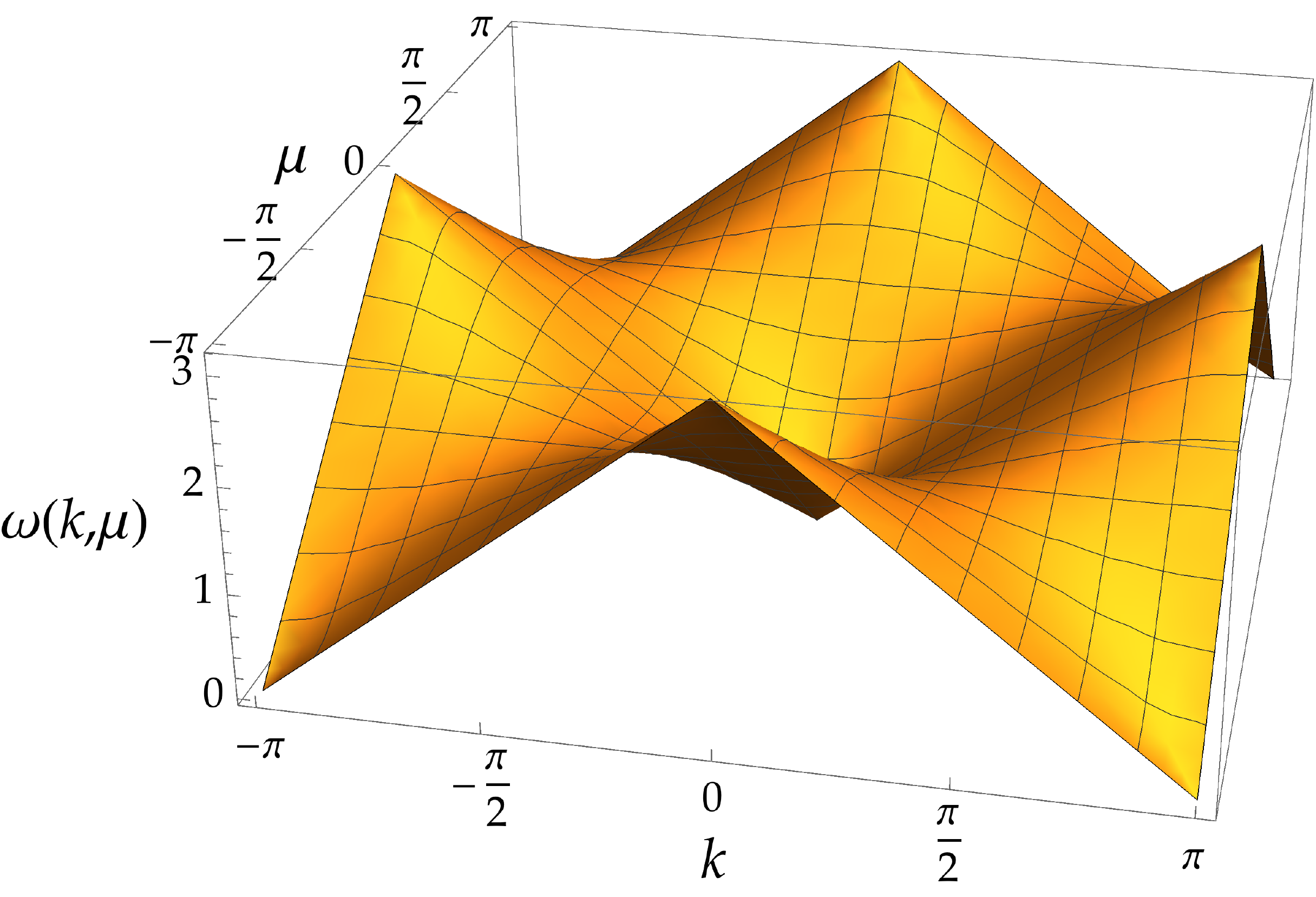}
\includegraphics[width=.32\textwidth]{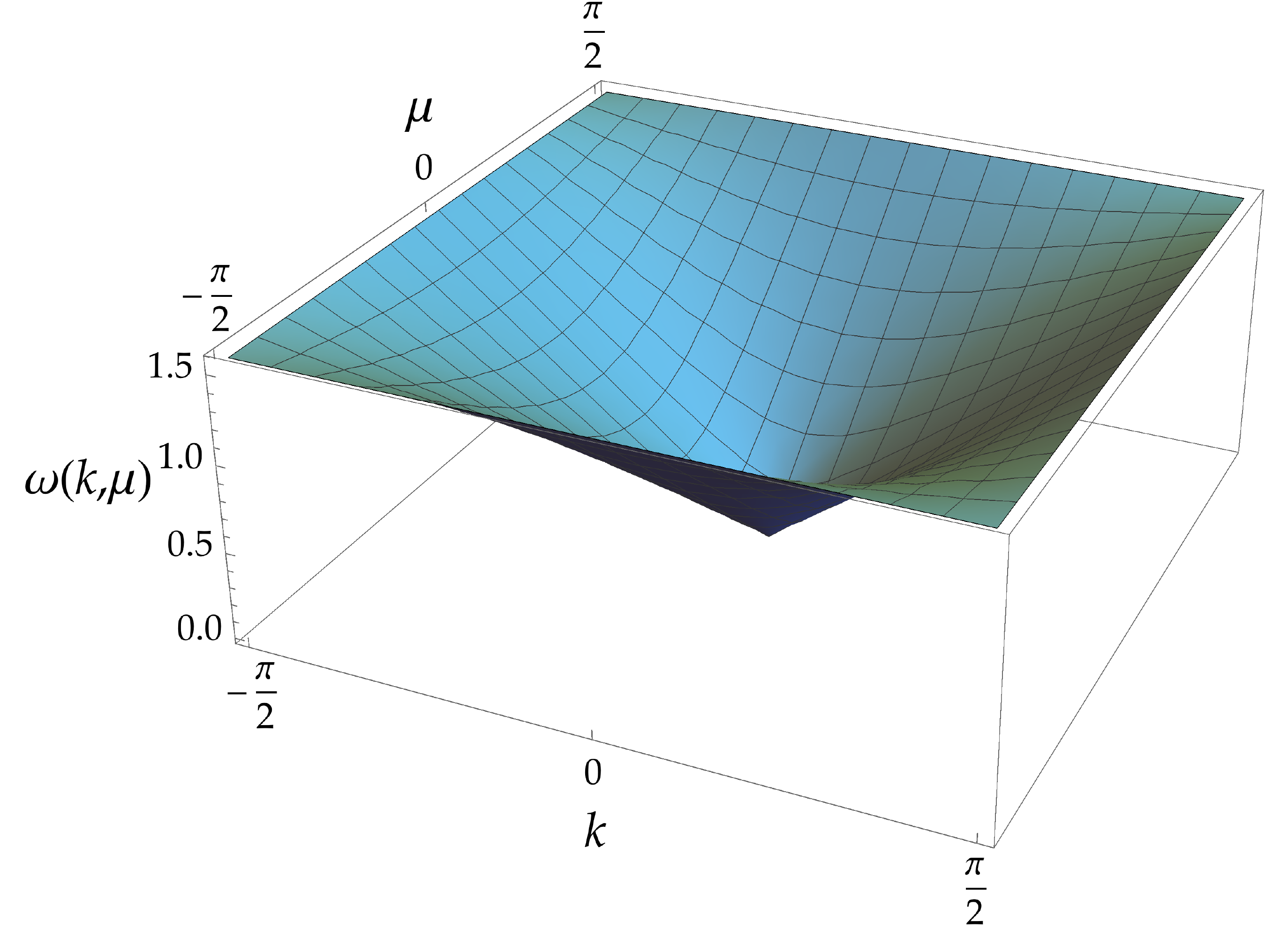}
\includegraphics[width=.32\textwidth]{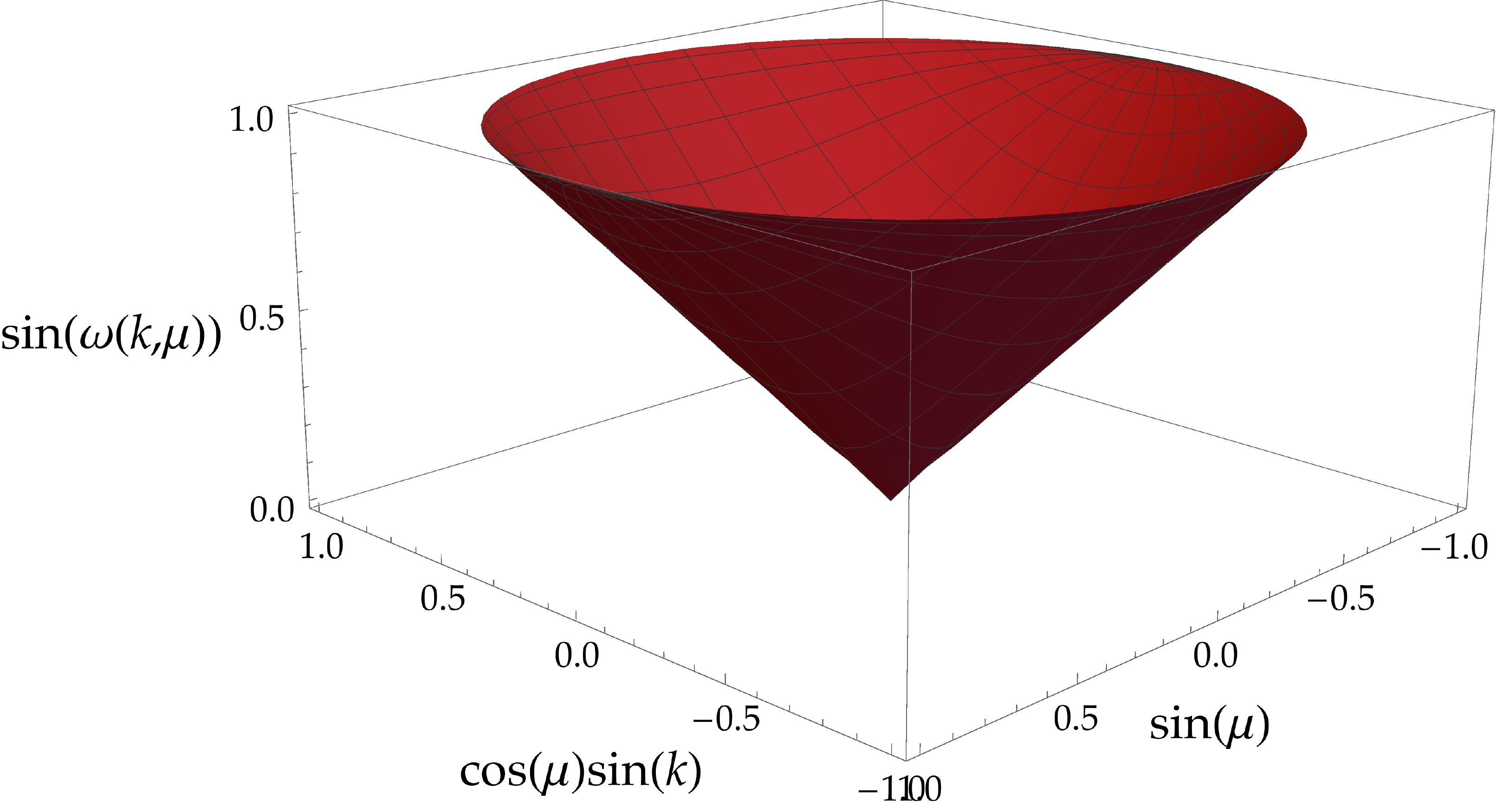}
\caption{\emph{Left}: Dispersion relation of the
  Dirac Quantum Walk with variable mass.
\emph{Middle}: The quantum walk mass shell $V$. The surface $V$ is the
graph of the dispersion relation restricted to the domain $B_0$.\\
\emph{Right}: The image $K:= n(V)$. The truncation is due to the 
condition $\bar n_\nu \bar n^\nu< 1$.}
\label{fig:cones}
\end{figure*}

 Therefore, for any $i\in\{0,1,2,3\}$ and $(\kmom,\mu) \in B_i$, if
 $n_\mu(\kmom,\mu) \tau^\mu \ket{\psi(\kmom,\mu)} =0 $, there exists
 $(\kmom',\mu') \in B_0$ such that $n_\mu(\kmom,\mu) = n_\mu(\kmom',\mu')$ and
 $\ket{\psi(\kmom,\mu)} = \ket{\psi(\kmom',\mu')}$.
We may understand the $B_i$
regions as kinematically equivalent sets, and the quantum walk
dynamics is completely specified by the solution of 
Equation~\eqref{egv} in any of the regions $B_i$.




\section{Change of inertial frame}
\label{sec:change-inert-frame}

As we saw in the previous section, the solution of the 
eigenvalue equation \eqref{egv} in one of the regions $B_i$,
which were
defined in Eq.~\eqref{eq:regionsBi}, completely characterizes the quantum 
walk dynamics.
We then require that a change of reference frame
leaves invariant the eigenvalue equation  \eqref{egv}
restricted to the domain $ B_0 $.
From now on, unless otherwise
specified, we will assume
  $(\kmom,\mu)\in B_0$
and consequently we
remove the restriction symbol $\cdot\vert_{B_0}$ from all the maps.
It is also convenient to introduce the notation
\begin{align*}
  k:= (
\omega,
\kmom,
\mu
).
\end{align*}
Let us now consider the map 
\begin{align*}
n:
k
\mapsto
\begin{pmatrix}
\sin\omega\\
\bar{ n}(\kmom,\mu)
\end{pmatrix}=
\begin{pmatrix}
\sin\omega\\
\cos\mu\sin\kmom\\
\sin\mu
\end{pmatrix}
\end{align*}
The map $n$ defines a diffeomorphism between
the quantum walk mass-shell
\begin{align*}
  V=\{k \; \vert \; \omega=\arccos(\cos\kmom\cos\mu)\},\\
\end{align*}
defined by condition~\eqref{eq:massshell} and the 
truncated cone
\begin{align*}
K:=\{(x,y,z)\mid x^2+y^2=z,0\leq z\leq1\}.
\end{align*}
both represented in Fig.~\ref{fig:cones}. In the following we will 
also use the non truncated null mass shell denoted by
\begin{equation}
K_0:=\{(x,y,z)\mid x^2+y^2=z\}.
\end{equation}

We are now ready to give a formal definition of change of reference frame.

\begin{definition}[Change of inertial reference frame]
\label{def:change-inert-frame}
  A change of inertial reference frame for the Dirac walk is a triple
  $(k', a, M )$ where:
 \begin{itemize}
 \item ${k}' :V\to V,\ k\mapsto k'(k)$, $k'\in\Diff(V)$ (the diffeomorphism group of the mass shell $V$),
\item $a \in C^{\infty}(V,\mathbb{C})$, $\ k \mapsto a(k)$ is a smooth complex  function,  
\item $M\in SL(2,\mathbb{C})$
 \end{itemize}
 such that:
  \begin{align}\label{eq:invarianceeigeneq2}
    \begin{aligned}
    &n_\mu(k) \tau^\mu \psi(k) = 0 \Leftrightarrow  n_\mu(k') \tau^\mu 
    \psi'(k') =0,  \\
    &\psi'(k') = e^{i a(k')}M \psi(k) \\
    \end{aligned}
  \end{align}
  for any $k\in V$.
  We denote with  $\mathcal{S}_{D}$ the group of changes of inertial reference
  frame (\emph{symmetry group} for short) for the
  Dirac   quantum walk with variable mass.
\end{definition}

According to Definition~\ref{def:change-inert-frame}, a change of
inertial frame is a relabeling $k'(k)$ of the constants of motion
of the quantum walk such that the eigenvalue equation is preserved in the region
$B_0$. The same definition straightforwardly generalises to the
other regions $B_i$.
The crucial assumption
in  Definition~\ref{def:change-inert-frame}
is that the linear transformation
$M$, which acts on the internal degrees of freedom, does not depend on
the value of $k$ \footnote{We could also have considered the case
$M \in GL(2,\mathbb{C})$. However, this choice would only have 
introduced a $k$-independent multiplicative constant. The symmetry
group would have been $\mathcal{S}'_D = \mathcal{S} \times
\mathbb{C}$, i.e. the symmetry group of the 
$M \in SL(2,\mathbb{C})$ case trivially extended by the direct product
with the multiplicative action of $\mathbb{C}$.}. Without this
limitation, the notion of symmetry group
would become trivial, allowing
any bijection between the set of solutions.
On the other hand, we presently lack a more physically grounded motivation for
this assumption.

From Definition \ref{def:change-inert-frame} the following rules
easily follow
\begin{align}
  \begin{aligned} \label{eq:rulesmulti}
&  [ (k', a, M )\psi ](k) = e^{i a(k)}M \psi(k'^{-1}(k))\\ 
& ( k'', b, N )\circ (k', a, M ) = (k''\circ k', b+a \circ k''^{-1}, NM ).  
  \end{aligned}
\end{align}

Let us now characterize the symmetry group $\mathcal{S}_D$.  Clearly,
the simplest example of change of inertial frame is the one given by
the trivial relabeling $k'=k$, the identity matrix $M=I$, and 
arbitrary $a(k)$.  Moreover, it is easy to realise that when $a(k)$ is
a real linear function, i.e. $a(k) = a_\mu k^{\mu}$ with $a_\mu \in \mathbb{R}^{3}$ , we recover the group
of translations in $3$ dimensions
(translations in the direction corresponding to the
varible $\tau$ conjugated to $\mu$ are also admissible).

We proceed with complete charaterization of the full symmetry group.
The basic result is the following lemma.
\begin{lemma}\label{lem:changeinert}
  Let $(k',a,M)$ be a change of inertial frame for the Dirac
  walk. Then we have
\begin{align}
  \label{eq:eigenpreserved}
  \begin{aligned}
  & f(k')  n_\mu(k')  =L^{\nu}_{\mu} n_{\nu}(k),\  \forall k\in V ,
  \end{aligned}
    \end{align}
  where $L\in SO^+(1,2)$, and 
  $f(k')$   is a suitable non null real function.
Moreover,
$M \in SL(2,\mathbb{R})$
such that
${{M}^{-1}}
 w_\mu \tau^\mu {M}  =  L_\mu^{\nu} w_\nu \tau^{\mu} $, for any $w\in{\mathbb R}^3$.
\end{lemma}
\Proof
Clearly we have that
$n_\mu(k) \tau^\mu \psi(k) = 0 \Leftrightarrow
      e^{i a(k')}n_\mu(k') \tau^\mu M
      \psi(k) =0 $
      for any $k\in V$
      if and only if
      $\sigma_2 n_\mu(k) \tau^\mu \psi(k) = 0 \Leftrightarrow
      {M^\dag} \sigma_2 n_\mu(k') \tau^\mu M
      \psi(k) =0 $, 
      because 
      $M \in SL(2,\mathbb{C})$.
From  Equation~\eqref{eq:projectorrank1}
we have that
$ \sigma_2 n_\mu(k) \tau^\mu $ is proportional to a  rank one projector
and therefore
we must have
\begin{align}
 & \begin{aligned}
g(k') n_\mu(k) \tau^\mu &=    \sigma_2    {M^\dag}
  \sigma_2 n_\mu(k') \tau^\mu M.
  \end{aligned}
\end{align}
Now, the right hand side is a linear combination of $I,\sigma_x,\sigma_y,\sigma_z$. 
By the above identity, however, we conclude that the right hand side must also be 
a linear combination of $\tau^\mu$ only. Thus
\begin{align}
     & \sigma_2    {M^\dag}
  \sigma_2 n_\mu(k') \tau^\mu M =:L_\mu^{\nu} n_\nu(k') \tau^{\mu} 
   \label{eq:equivop1}
\\
 &\implies  g(k') n_\mu(k)  =L_\mu^{\nu} n_\nu(k')
      \label{eq:equivop}  
\end{align}
for some non-null scalar function $g(k')$, and some linear map $L$. 
Then, since $M\in SL(2,\mathbb{C})$, 
we have that $L\in SO^+(1,2)$, and that $g(k')$ must be a real function.
Then $ M$ is a two dimensional representation of
$SO^+(1,2)$ which implies $ M \in SL(2,\mathbb{R})$.\qed

\begin{corollary}\label{lem:changeframe2}
  Let $(k',a,M)$ be a change of inertial frame for the Dirac walk.
  Then we have
  \begin{align}
    & L^{\nu}_{\mu} n_\nu(k) = \mathcal{D}_{f}  n_\mu(k')  \\
    &\begin{aligned}
    \mathcal{D}_f  :    \mathbb{R}^3 &\to \mathbb{R}^3,\ 
  n \mapsto f(n) n
  \label{eq:nonlinscal}
    \end{aligned} 
  \end{align}
  where $L \in SO^+(1,2)$ and $f:\mathbb R^3\to\mathbb R^3$ is a smooth function such that
  $\mathcal{D}_f$ is injective.
\end{corollary}
\Proof
Let $f(k')$ be as in Lemma~\ref{lem:changeinert}.
Since $n(k')$ is a diffeomorphism, we may consider $f$ as a function of
$n$, namely $f(n) := f(k'(n))$.
Let us now assume that $ \mathcal{D}_{f}$ is not
injective.
Then
we would have,
$
   \mathcal{D}_{f} \circ n(k'_1) = \mathcal{D}_{f} \circ n(k'_2) 
$
for some $k'_1 \neq k'_2$.
From Eq. \eqref{eq:eigenpreserved} we then have
$L_{\mu}^\nu n_\nu( k_1) = L_{\mu}^\nu n_\nu( k_2) $.
However, since both maps $k'(k)$ and $ L$
are invertible, this would imply $k_1=k_2$. \qed

We can finally prove the characterization of the symmetry group of the
Dirac walk.
\begin{proposition}\label{prop:changeframefinal}
  The triple  $(k',a,M)$ is a change of inertial frame for the Dirac walk
  if and only if
  \begin{align}
&    \label{eq:finalform}
    k'(k) =  [n^{-1} \circ {\mathcal{D}_{f}}^{-1} \circ L  \circ
                  \mathcal{D}_{g} \circ n ](k) \\
       \label{eq:finalM}
 &         M \in
    SL(2,\mathbb{R}), {{M}^{-1}}
  w_\mu \tau^\mu {M}  =  L_\mu^{\nu} w_\nu \tau^{\mu} ,\forall w\in\mathbb{R}^3\\
&  a\in C^\infty(V,\mathbb{C}),
  \end{align}
where $\mathcal{D}_{f}$ and $\mathcal{D}_{g}$ are two diffeomorphisms between $K$ and $K_0$, of the form of Eq.~\eqref{eq:nonlinscal}, and $L\in SO^+(1,2)$.
\end{proposition}
\Proof
From Corollary \ref{lem:changeframe2} we have that
\begin{align*}
  k'(k) =   [  n^{-1} \circ {\mathcal{D}_{f}}^{-1} \circ L \circ n ] (k),
\end{align*}
where $L\in SO^+(1,2)$, and $\mathcal{D}_f$ is of the form of Eq.~\eqref{eq:nonlinscal}.
Let now $\mathcal{D}_{g}$ be any
diffeomorphism  of the same form between $K$ and $K_0$.  Since $K$ is star shaped, such a $\mathcal{D}_{g}$ exists
(see Appendix \ref{invertible_f} for an example).
Then 
\begin{align*}
  k'(k) =   [  n^{-1} \circ {\mathcal{D}_{\bar{f}}}^{-1}
  \circ L  \circ \mathcal{D}_{g} \circ n ] (k)\\
  \mathcal{D}_{\bar{f}}^{-1} := {\mathcal{D}_{f}}^{-1} \circ L \circ \mathcal{D}_{g}^{-1} \circ L^{-1},
\end{align*}
where
$\mathcal{D}_{\bar{f}}$ is a diffeomorphism between $K$ and $K_0$.
\qed

\begin{figure*}[t]
  \includegraphics[width=0.8\textwidth]{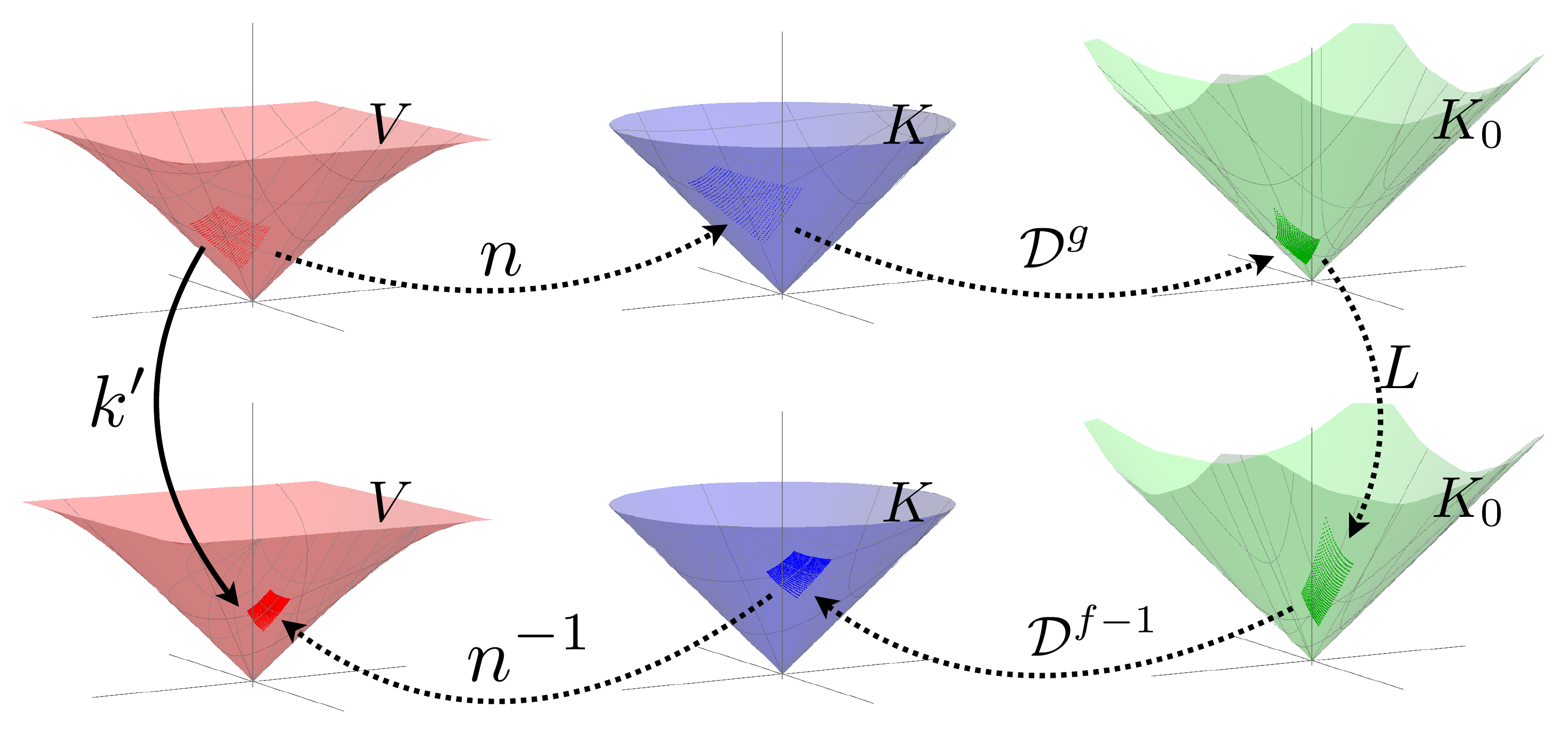}
\caption{Pictorial representation of a change of inertial frame for
  the Dirac Walk with variable mass. }
\label{fig:transf}
\end{figure*}

We provide a pictorial representation of the change of inertial frame in Fig. \ref{fig:transf}.

From Equation \eqref{eq:finalform}, it follows that the
diffeomorphisms $n \circ k' \circ n^{-1}$ form a subgroup
$ G$ of $\Diff(K)$ which is the product of a non linear
realization of $SO^+(1,2)$ and a group $M_K$ of nonlinear
dilations of $K$.
\begin{lemma}\label{lmm:dillorentz}
  Let $ G \subseteq \Diff(K)$ such that
  $\mathcal{G} \in G $ iff
  $\mathcal{G} = n \circ k' \circ n^{-1} $ where $k'$ obeys
  Eq.~\eqref{eq:finalform}. Then we have
\begin{align}
  \label{eq:decompogroup}
&  G = D_K \rtimes SO^+_f(1,2), \\
&  D_K:= \{  \mathcal{M} \in \Diff(K) \; | \; \mathcal{M}(v) =
  m(v) v \; \forall v\in K\},\\
&  SO^+_f(1,2) := \{   \mathcal{L}\in \Diff(K) \; | \;
  \mathcal{L} = \mathcal{D}_{f }^{-1} \circ L \circ \mathcal{D}_{f} \},
\end{align}
where $m(v)$ is a real function on $K$, $L \in SO^+(1,2) $
and
$\mathcal{D}_{f}: v \mapsto (1-(v_x^2+v_y^2))^{-1} v$.  
\end{lemma}
\Proof
Let us fix an arbitrary $\mathcal{G} \in G $.
From Eq. ~\eqref{eq:finalform}
we have that $\mathcal{G} =\mathcal{D}_{h}^{-1} \circ L  \circ
\mathcal{D}_{g} $, 
where $\mathcal{D}_{h}$ and $\mathcal{D}_{g}$ are two diffeomorphisms between $K$ and $K_0$, of the form of Eq.~\eqref{eq:nonlinscal},
and $L\in SO^+(1,2)$.
Let us define
\begin{align*}
&\mathcal{L} :=
\mathcal{D}_{f }^{-1} \circ L \circ \mathcal{D}_{f}  \\
 & \mathcal{M}:=\mathcal{L}^{-1}\circ \mathcal{G} =\mathcal{D}_{f}^{-1 } \circ L^{-1} \circ \mathcal{D}_{f} \circ
  \mathcal{D}_{h}^{-1} \circ L  \circ \mathcal{D}_{g}.
\end{align*}
One can verify that
$\mathcal{L}  \in SO^+_f(1,2) $ and $ \mathcal{M} \in D_K$.
Therefore, as $\mathcal G=\mathcal L\circ\mathcal M$,
we have $G=SO^+_f(1,2)D_K$.
Let now $\mathcal L\in SO^+_f(1,2)$ and $\mathcal M\in D_K$, and $\mathcal{L} =\mathcal{M}$. Then
\begin{align*}
L f(v) v = \mathcal D_f[m(v) v]=g(v)v ,\quad\forall v\in K,
\end{align*}
for some $g:K\to\mathbb R$. This implies that $L=I$, namely the intersection between $ SO^+_f(1,2) $ and $ D_K$
is only the identity map. 
Finally, 
$\mathcal{L} \circ \mathcal{M} \circ \mathcal{L}^{-1} \in D_K $ for
any $\mathcal{L} \in SO^+_f(1,2) $ and $\mathcal{M} \in D_K$,
i.e. $D_K$ is normal in $G$.\qed
We remark that the choice of
$\mathcal{D}_{f} $ in the definition of the subgroup $SO^+_f(1,2) $ is
arbitrary. One can choose any other diffeomorphism between $K$
and $K_0$ that satisfies Eq.~\eqref{eq:nonlinscal}.
The subscript $f$ in
$SO^+_f(1,2) $ is just a reminder that this group is a nonlinear
realization of $SO^+(1,2) $, i.e.~an homomorphism of
$SO^+(1,2) $ on $\Diff(V)$. Clearly, $SO^+(1,2) $  and
$SO^+_f(1,2) $
are isomorphic.

The decomposition of the symmetry group of the Dirac quantum walk with
variable mass is now easily provided by the following Proposition.
\begin{proposition}
  Let $\mathcal{S}_D$ be the symmetry group of
   Dirac quantum walk with
   variable mass.
   Then we have
   \begin{align}
     \label{eq:decompofinal}
     \mathcal{S}_D =   C^{\infty}(V,\mathbb{C})  \rtimes (D_K \rtimes SO^+(1,2))
   \end{align}
\end{proposition}
\Proof
The result follows from Lemma~\ref{lmm:dillorentz}
and Equation~\eqref{eq:rulesmulti}.
\qed

Thanks to the decomposition~\eqref{eq:decompofinal},
we easily see that the symmetry group $\mathcal{S}_D$
contains
  $\mathbb{R}^3 \rtimes SO^+(1,2)$
  as a subgroup, i.e. the Poincar\'e group in 2+1 dimensions,
  where the Lorentz transformations are nonlinearly deformed.
  Since one of the dimensions is
not spatial, but  associated with the mass parameter,
the subgroup $\mathbb{R}^3 \rtimes SO^+(1,2)$
is interpreted as a variation of the
de Sitter group, which occurs in 3+1 space-time dimensions.

Moreover, we remark that the subgroup given by
\begin{align}
  \label{eq:deflor}
  \begin{aligned}
 &    k'(k):= n^{-1}\circ \mathcal{D}_{f}^{-1} \circ L  \circ
  \mathcal{D}_{f} \circ n 
  \\
&  L =
\begin{pmatrix}
  \cosh(\xi) &\sinh(\xi)& 0 \\
  \sinh(\xi) &\cosh(\xi)& 0 \\
  0&0&1
\end{pmatrix},
  \end{aligned}
\end{align}
 is a non linear representation of the 1+1 dimensional Lorentz
group as the ones considered within the context of doubly special
relativity
\cite{amelino2001testable,magueijo2003generalized,PhysRevD.84.084010}.
If the Jacobian matrix  of the non linear map
$\mathcal{D}_{f}$ is the identity at the origin, it means
that the non linear Lorentz transformations recover the usual
linear ones in the limit of small wave-vectors.
This is the case for the non linear map given in Appendix~\ref{invertible_f}.

We have therefore characterized the full symmetry group of the Dirac
walk with variable mass, and we showed that in the small wave-vector limit it contains the usual Poincar\'e symmetry.
Now we can proceed giving an alternative definition of change of inertial frame,
starting from Definition~\ref{def:change-inert-frame} with the additional requirement
that the mass term is left unchanged.

\begin{definition}[Change of Inertial frame with fixed $\mu$]
\label{def:fixmu}
A change of inertial frame, which leaves unchanged the third component $\mu$, is a triple
 $(k', a, M )$ where
  \begin{align*}
	&{k}' :V\to V
	,\  {k} :=    
	\begin{pmatrix}
      	\omega\\
     	\kmom\\
	\mu 
    	\end{pmatrix}
	\mapsto
    	{k}'({k}) :=
    	\begin{pmatrix}
      	\omega'(k,\mu)\\
     	\kmom'(k,\mu)\\
	\mu 
      \end{pmatrix} \\
  \end{align*}
   is a diffeomorphism,
$a: V\to\mathbb C,\ k \mapsto a(k)$ is a smooth map,
 and $M \in SL(2,\mathbb{C})$ such that:
  \begin{align}\label{eq:invarianceeigeneq}
    \begin{aligned}
    &n_\mu(k) \tau^\mu \psi(k) = 0 \Leftrightarrow  n_\mu(k') \tau^\mu 
    \psi'(k') =0,  \\
    &\psi'(k') = e^{i a(k)}M \psi(k) ,\\
    \end{aligned}
  \end{align}
  for any $k\in V$.

\end{definition}

The analysis of Appendix~\ref{sec:symmetry-dirac-walk} allows one to show that starting from Definition \ref{def:fixmu}, the group of changes of inertial frame with fixed $\mu$ is characterized in terms of the group 
\begin{equation}
G\cong SO^+(1,1)\ltimes\mathbb Z_2
\end{equation}
generated by the matrices
\begin{align}
L=SDS^{-1},\ L_+=SFS^{-1},
\label{eq:generg}
\end{align}
with
\begin{align*}
S&=
\begin{pmatrix}
1& 1 & \sin\mu\\
-\cos\mu&\cos\mu&0\\
\sin\mu&\sin\mu&1
\end{pmatrix},
\end{align*}
and
\begin{align*}
D=\begin{pmatrix}
e^{-\beta} &0&0\\
0&e^{\beta}&0\\
0&0&1
\end{pmatrix}\ ,\ 
F=\begin{pmatrix}
0 &1&0\\
1&0&0\\
0&0&-1
\end{pmatrix}\ .
\end{align*}
\begin{proposition}\label{prop:changeframefixmu}
  The triple  $(k',a,M)$ is a change of inertial frame for the Dirac walk
  if and only if 
  \begin{align}
    \label{eq:finalformfixmu}
    k'(k) =  [n^{-1} \circ {\mathcal{D}_{f}}^{-1} \circ L  \circ  \mathcal{D}_{f} \circ n ](k)
  \end{align}
where 
\begin{align*}
\mathcal{D}_{f}(w):=\frac{\sin\mu}{T^3_\nu w^\nu}w, 
\end{align*}
where $T=\{T^\mu_\nu\}$, $T\in  SO^+(1,1)\ltimes\mathbb Z_2$.
\end{proposition}

As shown in Appendix \ref{sec:change-inert-frame}, the above group does not provide the expected phenomenology of a Lorenz group of boosts in 1+1 dimensions. This result then justifies the analysis of the full symmetry group $SO(1,2)$, starting from a definition of change of inertial frame which involves also $\mu$ as a dynamical degree of freedom. 

\section{Conclusion}
In this paper we derived the group of changes of inertial reference
frame for the Dirac walk in 1+1 dimension.  If the mass of the walk is
fixed, the group of admissible symmetries is inconsistent with the interpretation of the wave-vector as momentum. Therefore, we defined a Dirac walk
with variable mass, and studied the symmetry group of the latter.  As
a result, one finds a group of transformations that, along with
$\omega$ and $\kmom$, modify also the variable $\mu$, that defines the
mass term. Such a group can be considered as the 1+1-dimensional
counterpart of the de Sitter group, that acts on 
the walk mass shell by a realisation in terms of a group of
diffeomorphisms. Along with the de Sitter group one is forced to
consider a group of non-linear rescaling maps, so that the final group
is a semidirect product of these two components.

\acknowledgments This publication was made possible through the
support of a grant from the John Templeton Foundation under the
project ID\# 60609 Causal Quantum Structures. The opinions expressed
in this publication are those of the authors and do not necessarily
reflect the views of the John Templeton Foundation.

\bibliographystyle{apsrev4-1}
\bibliography{bibliography}

\appendix

\section{An example of a rescaling function}
\label{invertible_f}



We now provide an example of a real function $f$
such that the map $\mathcal D_f$ is a diffeomorphism between $K$ and $K_0$. 
In order to have $\mathcal D_f$ surjective, $f$ must be singular at
the superior border of the truncated cone.  Hence we define
\begin{align*}
&f: n\to\mathbb{R}\\
&f( n):=\frac{1}{1-n_2^2-n_3^2}.
\end{align*}
The latter function is manifestly singular at the border of $K$ and is monotonic versus $\| n\|_E$
($\|  \cdot \|_E$ is the Euclidean norm).
With this choice of $f$, it easy to very that
$\mathcal D_f$ is a diffeomorphism between $K$ and $K_0$.

\section{Symmetry of the Dirac Walk with fixed $\mu$}
\label{sec:symmetry-dirac-walk}
The analysis of the symmetry transformations of the Dirac Walk with fixed $\mu$ follow the same steps as in the case of variable mass .
The condition that the third component $\mu$ of the vector $k$ in the eigenvalue equation of the Dirac walk
is fixed,
implies that, for a  fixed value of $\mu$  and any
$\kmom \in (-\frac\pi 2, \frac \pi 2]$, we need to satisfy the system of equations
\begin{align}
  \left \lbrace
  \begin{aligned}
  n^\sigma(\kmom',\mu)& = \varphi(\kmom,\mu,L) L^{\sigma}_\nu n^\nu(\kmom,\mu)  \\
  n^3(\kmom',\mu) &= n^3(\kmom,\mu) = \sin\mu        
  \end{aligned}
  \right.
  \label{eq:fixmas}
\end{align}
where $L \in SO^+(1,2)$ and $\varphi$ is a non null function that may generally depend on $L$.
So the transformed $n$ is
\begin{align}
\label{fixedmu}
\frac{1}{\varphi(\kmom,\mu,L)}
\begin{pmatrix}
\sin{\omega(\kmom')}\\
\cos\mu\sin(\kmom')\\
\sin\mu
\end{pmatrix}
=L
\begin{pmatrix}
\sin{\omega(\kmom)}\\
\cos\mu\sin\kmom\\
\sin\mu
\end{pmatrix}.
\end{align}
Considering the equation for the third component, we can easily obtain a form for the dilation function, namely
\begin{align}
\label{fL}
\frac{1}{\varphi(\kmom,\mu,L)}=\frac{L^3_\nu{n}^\nu(\kmom,\mu)}{\sin\mu}.
\end{align}
Notice that the image of the map $n$ (for fixed value of $\mu$) is the hyperbolic arc given by the intersection  of $K$ and the plane of constant $\mu$. The extremal points $u,v$ correspond respectively to $\kmom=\pm\pi/2$, and are given by
\begin{equation}
u=(1, \cos\mu,\sin\mu), \quad v=(1,-\cos\mu,\sin\mu).
\end{equation}
Since the transformation $L$ is linear, it maps extremal points to extremal points, and from Eq. \eqref{fixedmu} we must have

\begin{align}
\label{eq:vee}
\left\{\begin{aligned}
Lu
&=\eta u
\\
Lv
&=\xi v
\end{aligned}\right.\quad\text{ or } \quad 
\left\{\begin{aligned}
Lu
&=\eta v
\\
Lv
&=\xi u
\end{aligned}\right. ,
\end{align}
with $\eta,\xi\in\mathbb{R}$.
We start focusing our attention on  the leftmost conditions in~\eqref{eq:vee}. 

At this point we can characterize  the subgroup starting from a complete set of eigenstates $\{u, v, w\}$. The vector $ w$ is such that
\begin{align*}
\forall a,b\in\mathbb{R},\;w_\nu(a u+b v)^\nu=0,
\end{align*} 
hence $ w=(\sin\mu, 0, 1)$. Moreover, it is an eigenvector of $L$, since 
\begin{align*}
0=L^\sigma_\nu  w_\sigma L^\nu_\tau(a u+b v)^\tau=L^\sigma_\nu  w_\sigma (\eta a u+\xi b v)^\nu,
\end{align*}
then $L^\sigma_\nu  w^\nu=\theta  w^\sigma$, for some real $\theta$.
Considering that $L\in SO^+(1,2)$, we have $\det L=1$, thus the product of the eigenvalues $\eta\theta\xi=1$. Moreover, 
\begin{align*}
L^\sigma_\nu v_\sigma L^\nu_\tau u^\tau=\eta \xi v_\nu u^\nu \implies \eta \xi=1,
\end{align*}
then $\theta=1$. Considering the  parametrization of $\eta=e^{\beta}, \xi=e^{-\beta}$, where $\beta\in \mathbb{R}$, 
we can diagonalize $L$ as 
\begin{align}
D&:=S^{-1}LS=
\begin{pmatrix}
e^{-\beta} &0&0\\
0&e^{\beta}&0\\
0&0&1
\end{pmatrix}\ ,\label{eq:elfixmu}\\
S&=
\begin{pmatrix}
1& 1 & \sin\mu\\
-\cos\mu&\cos\mu&0\\
\sin\mu&\sin\mu&1
\end{pmatrix}.\nonumber
\end{align}
Let us now consider the alternative transformations, defined by the rightmost
condition in~\eqref{eq:vee}. Repeating a similar analysis as before, we recover the two following transformations
\begin{align}
\label{nm}
N_{\pm}=
\begin{pmatrix}
0&\pm e^{\beta}&0\\
\pm e^{-\beta}&0&0\\
0&0&-1
\end{pmatrix},\ 
O_{\pm}=
\begin{pmatrix}
0&\pm e^{\beta}&0\\
\mp e^{-\beta}&0&0\\
0&0&1\\
\end{pmatrix}
\end{align}
Computing the square of transformations on the right we obtain
\begin{align*}
O_\pm^2=
\begin{pmatrix}
-1&0&0\\
0&-1&0\\
0&0&1
\end{pmatrix},
\end{align*}
then representing $O_\pm^2$ in the canonical basis, we have
\begin{align*}
T_\pm^2=&SO_\pm^2S^{-1},\\
T_\pm^2=&\begin{pmatrix}
-\frac{3-\cos(2\mu)}{2\cos^2\mu}&0&\frac{2\sin\mu}{\cos^2\mu}\\
0&-1&0\\
\frac{2\sin\mu}{\cos^2\mu}&0&\frac{3-\cos(2\mu)}{2\cos^2\mu}
\end{pmatrix}.
\end{align*}
We easily note that the following inequality holds
\begin{align*}
-\frac{3-\cos(2\mu)}{2\cos^2\mu}<0\quad \forall \mu
\end{align*}
namely the orthochronicity condition is not verified,  then $T_\pm^2\notin SO^+(1,2)$. Hence we are left with the transformations $N_\pm$ in~\eqref{nm}. Their representation
in the canonical basis is $L_{\pm}=SN_{\pm}S^{-1}$.
By explicit calculation, we see that $(L_{\pm})_1^1=\pm\sec^2\mu\cosh\beta+\tan^2\mu $. We can then exclude the transformation $L_-$, since it is manifestly not  orthochronous. Moreover, it is clear  that the transformations  can be obtained  as follows
\begin{align*}
L_+=LSFS^{-1},\qquad F=
\begin{pmatrix}
0&1&0\\
1&0&0\\
0&0&-1
\end{pmatrix},
\end{align*}
for some $L$. Therefore the allowed subgroup is $SO^+(1,1)\rtimes\mathbb Z_2$, where $SO^+(1,1)$ is the group of matrices $L$ in Eq.~\eqref{eq:elfixmu}. 

Considering the rescaling in Eq.~\eqref{fL},
we obtain the following expression for the changes of inertial frame
\begin{align}
k'(k)=(n^{-1}\circ\mathcal{D}_\varphi^{-1}\circ L\circ\mathcal{D}_\varphi\circ n)(k),
\label{eq:ell}
\end{align}
with $L\in SO^+(1,1)\rtimes\mathbb Z_2$.
At this point we want to study the resulting group in the relativistic regime, for small values of the mass parameter $\mu$. Deriving the expressions in Eq.~\eqref{eq:ell} with respect to $\beta$ in $\beta=0$, and expanding the generators to the first order in $\mu$, we obtain the following group generators
\begin{align*}
&\tilde J:=\varphi(\kmom,\mu,0)
\begin{pmatrix}
0&1&0\\
1&0&-\mu\\
0&\mu&0
\end{pmatrix}+\partial_\mu\varphi(\kmom,\mu,0)I.
\end{align*}
It is thus clear  that we do not recover the Lorentz group in $1+1$ dimension.

\end{document}